\documentclass[aps,prb,reprint,superscriptaddress,amsmath, showpacs]{revtex4-1}
\usepackage{graphicx, dcolumn}
\begin{document}
\title{Structure and Giant Inverse Magnetocaloric Effect of Epitaxial Ni-Co-Mn-Al Films}
\author{N. Teichert}\email{nteichert@physik.uni-bielefeld.de}
\affiliation{Center for Spinelectronic Materials and Devices, Department of Physics, Bielefeld University, 33615 Bielefeld, Germany}

\author{D. Kucza}
\affiliation{Center for Spinelectronic Materials and Devices, Department of Physics, Bielefeld University, 33615 Bielefeld, Germany}

\author{O. Yildirim}
\affiliation{Institute of  Ion Beam Physics and Materials Research, Helmholtz-Zentrum Dresden-Rossendorf e.V., 01328 Dresden, Germany}
\affiliation{Dresden University of Technology, 01062 Dresden, Germany}

\author{E. Yuzuak}
\affiliation{Ankara University, Department of Engineering Physics, Faculty of Engineering, 06100 Besevler, Ankara, Turkey}
\affiliation{Department of Nanotechnology Engineering, Faculty of Engineering, Recep Tayyip Erdogan University, 53100 Rize, Turkey}

\author{I. Dincer}
\affiliation{Ankara University, Department of Engineering Physics, Faculty of Engineering, 06100 Besevler, Ankara, Turkey}

\author{A. Behler}
\affiliation{IFW Dresden, Institute for Complex Materials, P.O. Box 27 01 16, 01171 Dresden, Germany}

\author{B. Weise}
\affiliation{IFW Dresden, Institute for Complex Materials, P.O. Box 27 01 16, 01171 Dresden, Germany}

\author{L. Helmich}
\affiliation{Center for Spinelectronic Materials and Devices, Department of Physics, Bielefeld University, 33615 Bielefeld, Germany}

\author{A. Boehnke}
\affiliation{Center for Spinelectronic Materials and Devices, Department of Physics, Bielefeld University, 33615 Bielefeld, Germany}

\author{S. Klimova}
\affiliation{Center for Spinelectronic Materials and Devices, Department of Physics, Bielefeld University, 33615 Bielefeld, Germany}
\affiliation{Department of Nano- and Biomedical Technology, Saratov State University, 410012 Saratov, Russia}

\author{A. Waske}
\affiliation{IFW Dresden, Institute for Complex Materials, P.O. Box 27 01 16, 01171 Dresden, Germany}

\author{Y. Elerman}
\affiliation{Ankara University, Department of Engineering Physics, Faculty of Engineering, 06100 Besevler, Ankara, Turkey}

\author{A. H\"utten} 
\affiliation{Center for Spinelectronic Materials and Devices, Department of Physics, Bielefeld University, 33615 Bielefeld, Germany}

\date{\today}
\begin{abstract}
The structural, magnetic, and magnetocaloric properties of epitaxial Ni-Co-Mn-Al thin films with different compositions have been studied. The films were deposited on MgO(001) substrates by co-sputtering on heated substrates. All films show a martensitic transformation, where the transformation temperatures are strongly dependent on the composition. The structure of the martensite phase is shown to be 14M. The metamagnetic martensitic transformation occurs from strongly ferromagnetic austenite to weakly magnetic martensite. The structural properties of the films were investigated by atomic force microscopy and temperature dependent X-ray diffraction. Magnetic and magnetocaloric properties were analyzed using temperature dependent and isothermal magnetization measurements.  We find that Ni$_{41}$Co$_{10.4}$Mn$_{34.8}$Al$_{13.8}$ films show giant inverse magnetocaloric effects with magnetic entropy change of 17.5\,J\,kg$^{-1}$K$^{-1}$ for $\mu_0 \Delta H=5\,\text{T}$.

\end{abstract}
\maketitle
\section{Introduction}
In the ongoing search for magnetocaloric materials, Mn-rich Heusler compound based ferromagnetic shape memory alloys (FSMA) of the system Ni-Mn-Z (Z=Sb, Ga, In, Sn) turned out to be very promising due to low cost of the containing elements and sizable magnetocaloric effects (MCE).\cite{Planes2009, Buchelnikov2012, Khan2007} Substitution of Co for Ni in Ni-Mn-Z is known to improve the metamagnetic behavior of the martensitic transformation, and thus the magnetocaloric properties as it increases the austenite Curie temperature $T_\text{C}^{\text{A}}$ and leads to a transformation from weakly magnetic martensite to ferromagnetic austenite.\cite{Han2008, Fabbrici2011, Kainuma2006nature, Kainuma2006apl, Kainuma2008}
The origin for the large change of magnetization at the transformation temperature is a change of the magnetic coupling between Mn atoms on Mn sites and Mn atoms on Z sites due to the change of the lattice constants.\cite{Aksoy2009, Behler2013}

Off-stoichiometric Ni-Mn-Al also shows a martensitic transformation but accompanied by only small changes of the magnetization and hence negligible MCE.\cite{Morito1996, Kainuma1996}  The compound crystallizes in a \textit{B}2+\textit{L}2$_{1}$ mixed phase where the \textit{B}2 phase is antiferromagnetic and the \textit{L}2$_{1}$ phase is ferromagnetic.\cite{Fujita2000, Acet2002, Manosa2004} Substitution of up to 10\,at.\% Co for Ni strongly promotes the ferromagnetism in the austenite phase and leads to a metamagnetic martensitic transformation.\cite{Kainuma2008}
The magnetization difference between austenite and martensite enables magnetic field induced reverse transformation together with an inverse magnetocaloric effect.\cite{Xu2010, Xuan2014, Kim2014}

Our interest is in epitaxial thin films of magnetocaloric materials as they present a good model system to study underlying physics due to the fixed crystallographic orientation. Additionally, thin films offer a high surface to volume ratio and if they are released from the substrate also ductility,\cite{Backen2010} and thus are promising for small scale magnetocaloric applications. In earlier studies we could show that the characteristics of the martensitic transformation and magnetocaloric properties of 200\,nm Ni-Mn-Sn thin films are comparable to those of bulk material,\cite{Auge2012, Yuzuak2013} so this film thickness is also chosen for the present work. Reports about Ni-Co-Mn-Al are very sparse in literature\cite{Mitsui2010, Rios2013}, and thus we want to give insight into the structural and magnetocaloric properties of epitaxial Ni-Co-Mn-Al thin films. Therefore, we prepared a set of films with different compositions and hence, different transformation temperatures.

\section{Experimental Details}
Three films with thickness of 200\,nm and different compositions were prepared, where the composition change is mainly in the Al content.  The films are listed in Tab.~\ref{tab:samples} and labeled after their Al content. 
The films were grown on MgO[001] substrates by magnetron co-sputtering in an ultra high vacuum system with base pressure better than $5\times 10^{-9}$\,mbar. The films were deposited from elemental Ni, Co, Mn, and Al targets. Before deposition of the Heusler compound the substrate was heated to 500$^{\circ}$C and a 35\,nm thick V seed-layer was deposited. During the subsequent deposition of the Heusler layer the substrate was rotated at 10\,rpm. All films are capped by a protective 2\,nm  MgO layer deposited by e-beam evaporation. The V seed layer can also act as a sacrificial layer as it can be removed by chemical wet-etching in order to obtain freestanding films. Investigation of the magnetocaloric properties of freestanding Ni-Co-Mn-Al films will be the subject of future studies.

\begin{table}
\caption{\label{tab:samples}List of the investigated thin films. The compositions are given in at.\%. $T_{\text{M}}$ and $T_{\text{A}}$ denote the martensitic and austenitic transformation temperatures, and $T_{\text{C}}$ the austenite Curie temperature as determined from magnetization measurements in section \ref{sec:magn}.}
\centering
\begin{tabular}[t]{cccccccc}
sample &  Ni& Co& Mn& Al& $T_{\text{M}}$ (K)& $T_{\text{A}}$ (K)& $T_{\text{C}}$ (K)\\ 
\hline
Al-14.3 & 40.7& 10.4& 34.6& 14.3&206&323&409\\
Al-13.8 & 41.0& 10.4& 34.8& 13.8& 348& 389 &416\\
Al-12.7 & 41.5& 10.6& 35.2& 12.7& 418& - &424\\
\end{tabular}
\end{table}

The film thickness was determined by X-ray reflectivity (XRR). Structural analysis was done by temperature dependent X-ray diffraction (XRD) measurements in the temperature range between 200\,K and 350\,K using Bragg Brentano optics with Cu K$_{\alpha}$ radiation and a custom built LN$_{2}$ cryostat. The surface morphology of the martensitic films was investigated by atomic force microscopy (AFM) at room temperature. Temperature and magnetic field dependent magnetic properties of Ni-Co-Mn-Al films were investigated with a superconducting interference device (SQUID, Quantum Design MPMS XL 7) in the temperature range from 50\,K to 400\,K and a vibrating sample magnetometer (VSM) employing either a heating (300\,K - 600\,K) or a cooling system (50\,K to 390\,K), where in-plane external field was applied.

\section{Results and Discussion}
\subsection{Structure}

\begin{figure}
\centering
\includegraphics[width=8 cm]{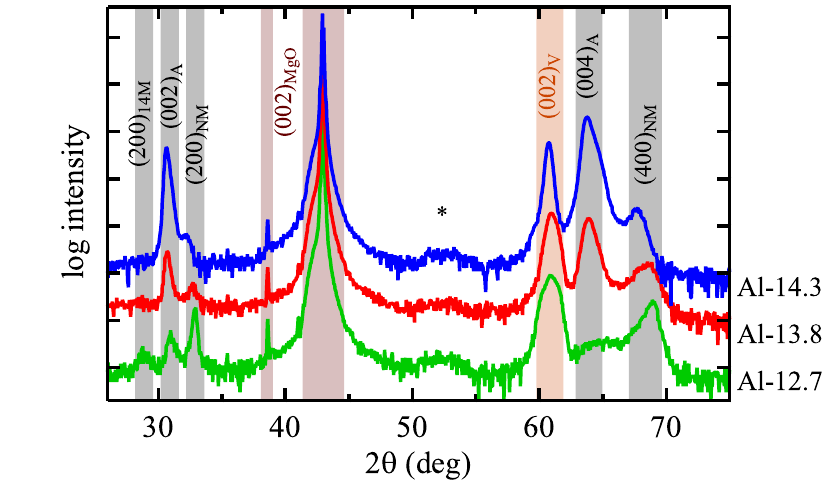}
\caption{\label{figure:xrd}Room temperature XRD patterns of the investigated epitaxial Ni-Co-Mn-Al films.}
\end{figure}
Figure~\ref{figure:xrd} shows the XRD patterns of all analyzed films at room temperature (RT). The films are grown epitaxially with the relation MgO[100](001)$\Vert$Ni-Co-Mn-Al[110](001). At 64$^{\circ}$ the (004)$_{\text{A}}$ peak of the cubic austenite of Ni-Co-Mn-Al is visible. The existence of the (002)$_{\text{A}}$ superstructure peak at 30.5$^{\circ}$ indicates \textit{B}2 structure. Odd superlattice reflections belonging to \textit{L}2$_{1}$ structure (e.g. (111)) were not found  within further analysis using a four-circle goniometer. This is in accordance to other studies of bulk Ni-(Co-)Mn-Al, where \textit{B}2 is the dominating structure.\cite{Acet2002, Kainuma2008, Mitsui2010} Nevertheless, we give the lattice constants with reference to \textit{L}2$_1$ for comparability to other Heusler compound based FSMAs. The (400)$_{\text{NM}}$ peak at 69$^{\circ}$ belongs to the martensite phase. From (004)$_{\text{A}}$ and (400)$_{\text{NM}}$ peak intensities at room temperature it is visible that the amount of martensite at RT and thus the transformation temperature increases with decreasing Al content which is also reported for Ni-Mn-Al.\cite{Kainuma1996} Further explanation of the indexing of the martensite reflections is found below. 

Besides  peaks belonging to Ni-Co-Mn-Al there are also the MgO (002) peaks from the substrate at 42.9$^{\circ}$ (Cu K$_{\alpha}$) and 38.6$^{\circ}$ (Cu K$_{\beta}$) and the (002) peak of the V buffer layer at 61$^{\circ}$ visible. The weak reflection marked by an asterisk is present in all films but could not doubtlessly be indexed. It probably belongs to a binary impurity phase.

\begin{figure}
\centering
\includegraphics[width=8.5 cm]{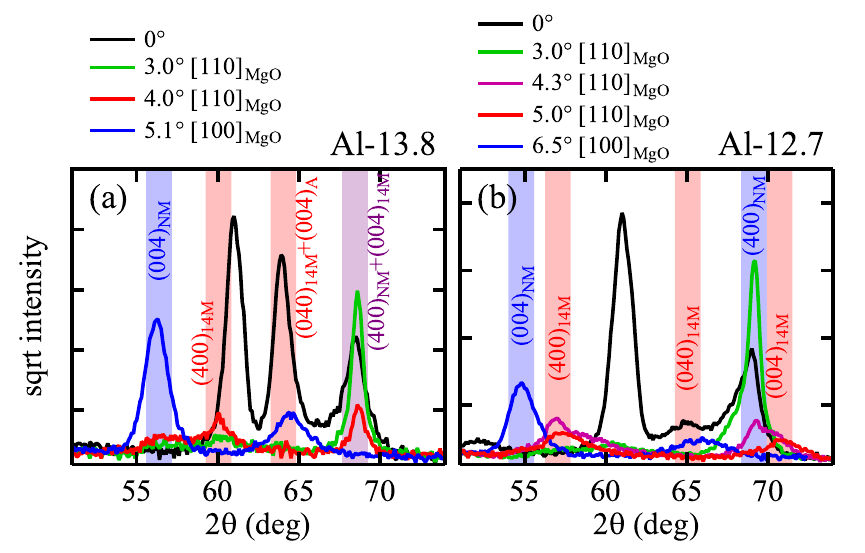}
\caption{\label{figure:tiltedxrd} XRD patterns at different tilt angles of the MgO substrate around its [110] and [100] direction for a) Al-13.8 and b) Al-12.7.}
\end{figure}

The films Al-13.8 and Al-12.7 which are mainly martensitic at room temperature allow for detailed investigation of the martensitic phase.
It is known that the martensite unit cells are tilted by small angles away from the substrate normal in order to build an almost exact habit plane.\cite{Kaufmann2011} This tilts make it necessary to adjust the sample alignment in the XRD system. This was achieved by measuring XRD patterns under certain tilt angles of the sample around the [100]$_{\text{MgO}}$ or [110]$_{\text{MgO}}$ direction. 
The results are depicted in Fig.~\ref{figure:tiltedxrd} and the peak positions allow identification of the martensitic phase as a 14\textit{M} modulated structure for both films. The peaks are indexed with respect to the \textit{L}2$_{1}$ unit cell.
The observed 14\textit{M} phase is proposed to be an adaptive phase constructed from tetragonal building blocks in (5$\bar2$)$_{2}$ periodicity in order to obtain an almost exact interface to the austenite.\cite{Khachaturyan1991, Pons2000, Kaufmann2010} The Bragg reflections of the tetragonal non-modulated (NM) variants are also visible in the XRD patterns. The lattice constants extracted from these XRD patterns are listed in Tab.~\ref{tab:latticeparameters}. 

The concept of adaptive martensite implies the following relations between 14\textit{M}, NM, and austenite: i) $b_{14M}=a_{\text{A}}$, ii) $c_{14M}=a_{\text{NM}}$, and iii) $a_{14M}=c_{\text{NM}}+a_{\text{NM}}-a_{\text{A}}$.\cite{Kaufmann2010} All these relations are almost exactly fulfilled. 

\begin{figure}
\centering
\includegraphics[width=8.5 cm]{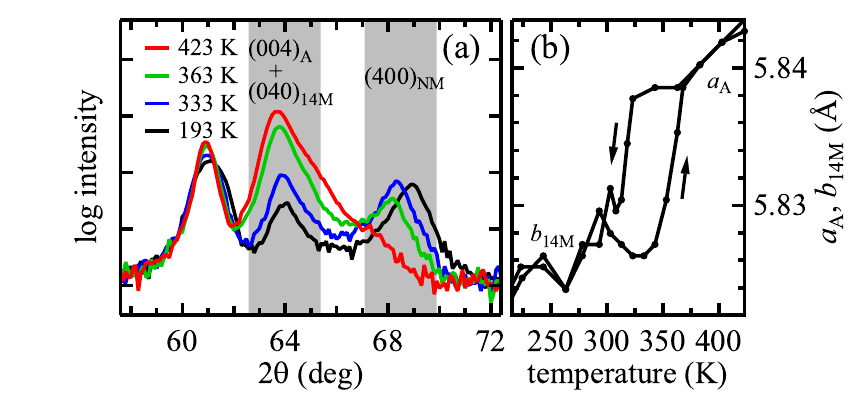}
\caption{\label{figure:tempxrd} a) XRD patterns of Al-13.8 at different temperatures during heating at zero tilt of the substrate. b) Temperature dependence of $a_{\text{A}}$ and $b_{14M}$. The arrows indicate the direction of temperature change.}
\end{figure}

The first relation concerns the peak at about 64$^{\circ}$. In order to distinguish $a_\text{A}$ from $b_{14M}$ the temperature dependence of this lattice parameter is analyzed. Therefore, XRD patterns were taken at different temperatures as shown in Fig.~\ref{figure:tempxrd}(a) for Al-13.8. Apart from the change of the peak intensities due to the martensitic transformation also changes of the peak positions are visible. The lattice constant related to the peak at about 64$^{\circ}$ is depicted in Fig.~\ref{figure:tempxrd}(b) and corresponds to $a_{\text{A}}$ at high temperatures and $b_{14M}$ below the martensitic transformation. The hysteresis in the temperature range of the martensitic transformation reveals a difference between $a_\text{A}$ and $b_{14M}$ of about 0.01\,\AA. 
Furthermore, from Fig.~\ref{figure:tiltedxrd}(a) it is visible that the (040)$_{\text{14\textit{M}}}$ appears not only at zero tilt but also at 5.1$^\circ$ tilt around [100]$_{\text{MgO}}$ at a slightly different angle. This peak belongs to a different unit cell orientation and shows a lattice constant that is larger by $0.04\,\text{\AA}$. 

The second relation seems to be exactly fulfilled for Al-13.8 since the (400)$_{\text{NM}}$ and (004)$_{\text{14\textit{M}}}$ can not be distinguished. For Al-12.7, however, those peaks can be distinguished and the lattice constants differ by $0.11\,\text{\AA}$. 

The third relation can easily be checked using the lattice parameters from Tab.~\ref{tab:latticeparameters} and fits also almost exactly. However, this analysis reveals slight differences between the ideal model of adaptive martensite and the measured unit cells and also slightly different lattice constants depending on the orientation of the unit cell. The reason for that can be an incommensurate 14\textit{M} microstructure. The decisive parameter for that is the twinning periodicity $d_{1}/d_{2}=(c_{14M}-a_{\text{A}})/(a_{\text{A}}-a_{14M})$, which is $d_{1}/d_{2}=2/5=0.4$ for a commensurate 14\textit{M} structure. The calculated values are 0.49 for Al-13.8 and 0.33 for Al 12.7, thus the microstructure is incommensurate. This results in a high density of stacking faults, which can be the reason why the mentioned relations i)-iii) are not exactly fulfilled.

\begin{table*}
\caption{\label{tab:latticeparameters}Lattice parameters of selected films given first in Heusler notation for comparison with epitaxial Heusler thin films from other studies and then in bct notation for comparison with bulk Ni-Mn-Al. All lattice constants are given in \AA. $\beta$ is the monoclinic angle and given in degrees.}
\centering
\begin{tabular}[t]{c|ccc|cccc|cccc}
sample &$a_{\text{NM}}$ & $c_{\text{NM}}$& $(c/a)_{\text{NM}}$ & $a_{\text{14M}}$ & $a_{\text{A}},b_{\text{14M}}$ &$c_{\text{14M}}$ & $(c/a)_{\text{14M}}$ &$a_{\text{14M}}^{\text{bct}}$&$b_{\text{14M}}^{\text{bct}}$&$c_{\text{14M}}^{\text{bct}}$&$\beta$ \\ 
\hline
Al-14.3 &  & && & 5.84&&&&& \\
Al-13.8 & 5.47  & 6.55&1.20 & 6.17 & 5.83 & 5.47& 0.89& 4.27& 29.5 & 2.74 & 94.4\\
Al-12.7 & 5.43 & 6.70&1.23& 6.47& 5.75& 5.32&0.82& 4.31 & 29.7 & 2.72 & 95.2\\
Ni-Co-Mn-In\footnotemark[1] &5.48&6.88&1.26&6.33&5.94&5.56&0.88\\
Ni-Mn-Ga\footnotemark[2]    &5.42&6.65& 1.23&6.18&5.78&5.62&0.91\\
Ni$_{50}$Mn$_{34}$Al$_{16}$\footnotemark[3]&&&&&&&&4.31&29.6&2.71&94.5\\
Ni$_{45}$Mn$_{40}$Al$_{15}$\footnotemark[3]&&&&&&&&4.34&29.7&2.71&94.7\\
\end{tabular}
\footnotetext[1]{Reference [\onlinecite{Niemann2010}].}
\footnotetext[2]{Reference [\onlinecite{Kaufmann2010}].}
\footnotetext[3]{Reference [\onlinecite{Kainuma1996}].}
\end{table*}

\begin{figure}
\centering
\includegraphics[width=8.5 cm]{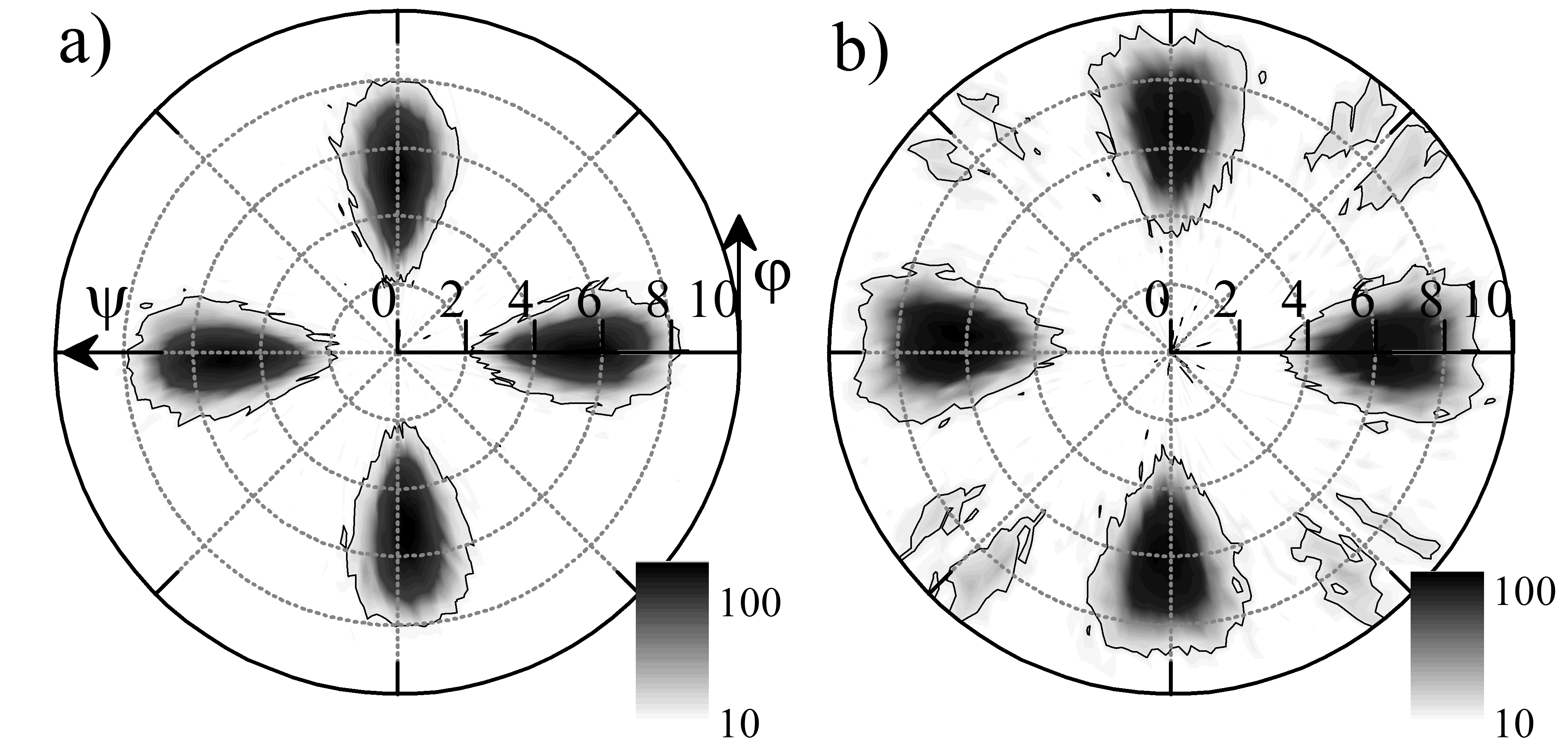}
\caption{\label{figure:polefigures} XRD pole figures of the (004)$_{\text{NM}}$ peaks for a) Al-13.8 and b) Al-12.7.}
\end{figure}

Figure~\ref{figure:polefigures} shows the pole figure measurements for the (004)$_{\text{NM}}$ peaks. It is visible that the main reflections of this peak are at $\varphi=0^\circ$, and $\psi\approx 5^\circ$ and $6.5^\circ$ for Al-13.8 and Al-12.7, respectively where $\varphi=0^\circ$ is equivalent to the [100]$_{\text{MgO}}$ direction. So, the main reflections of the pole figures fit to the used tilt angles in order to obtain maximum peak intensity. The observed tilt angles, and thus the orientation of the NM unit cell originates from the orientation of the NM cells inside the 14\textit{M} unit cell and the orientation of the 14\textit{M} unit cell with respect to the austenite. The tilt between 14\textit{M} and austenite is $\gamma=45^\circ-\arctan(c_{\text{14\textit{M}}}/a_{\text{14\textit{M}}})$ using the approximation that the 14\textit{M} unit cell is orthorhombic.\cite{Thomas2008} 
This results in $\gamma=3.42^\circ$ for Al-13.8 and $\gamma=5.55^\circ$ for Al-12.7. $\gamma$ describes a tilt of the 14\textit{M} unit cell around $b_{14M}$.\cite{Kaufmann2010}
The relevant NM unit cells inside the 14\textit{M} cell are inclined by $3.31^\circ$ and $3.93^\circ$ around $c_{14M}$ for Al-13.8 and Al-12.7, respectively. These tilt angles can be determined from the structure of the 14\textit{M} unit cell by basic geometry as described in Ref. [\onlinecite{Kaufmann2010}].
Combining these two tilts one can calculate the expected peak positions in the pole figures. The result is $\psi=4.7^\circ$, $\varphi=\pm 1^\circ$ for Al-13.8 and $\psi=6.8^\circ$, $\varphi=\pm 10^\circ$ for Al-12.7. The calculated $\psi$ angles fit almost precisely to the measured angles for both films. The larger calculated $\varphi$ for Al-12.7 also explains the broadening in $\varphi$ direction of the measured major peaks in the pole figure in Fig.~\ref{figure:polefigures}(b), which look like superpositions of two peaks at slightly different $\varphi$.
Due to 4-fold symmetry induced by the substrate, the according reflections in the other quadrants of the pole figures are also explained. 
The minor reflections in the (004)$_{\text{NM}}$ pole figure of Al-12.7 at $\psi=8^\circ$ and $\varphi=45^\circ \pm 7^\circ$ probably originate from a different orientation of the 14\textit{M} unit cell which is not present in Al-13.8. 

\begin{figure}
\centering
\includegraphics[width=8.5 cm]{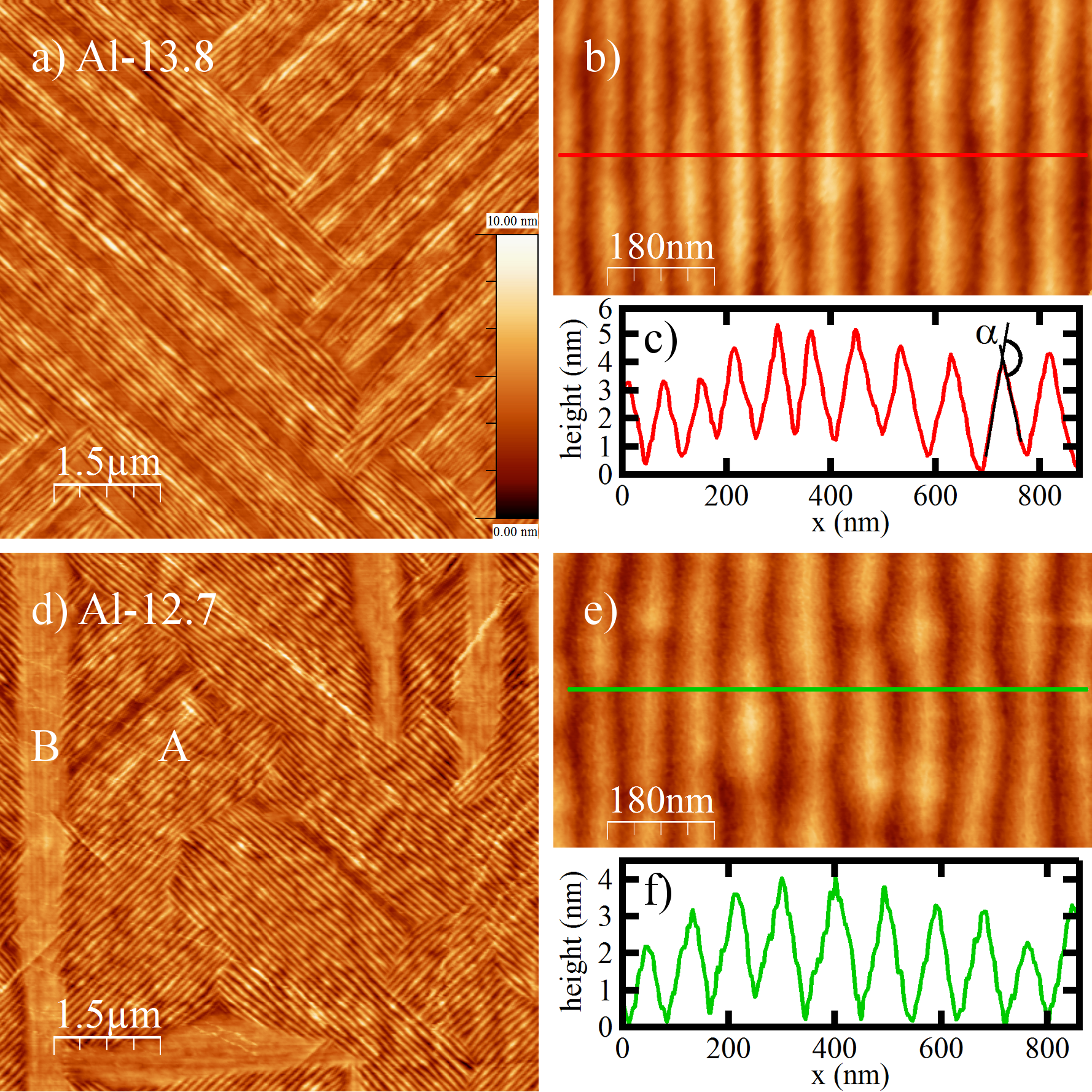}
\caption{\label{figure:afm} AFM micrographs and height profiles of Al-13.8 (a-c) and Al-12.7 (d-f). The picture margins of a) and d) are parallel to the substrate edges. The letters A and B label the two topography types found for Al-12.7.
 b) and e) show 45$^{\circ}$ rotated close-ups of the nanostructure and show the trace of the corresponding height profile, shown in c) and f), respectively.}
\end{figure}

In order to visualize the microstructure and to support the XRD results, the surface morphology of the films was analyzed by AFM. The micrographs for Al-13.8 and Al-12.7 are shown in Fig.~\ref{figure:afm}. The surface of Al-13.8 indicates an austenite/martensite mixed phase. About 80\% of the surface shows a typical martensitic microstructure with traces inclined by 45$^\circ$ to the substrate edges, which is also seen for epitaxial Ni-Mn-Ga thin films.\cite{Kaufmann2011} The periodicity of the variant traces is 83\,nm. The flat ribbons parallel to the twinning traces belong to the austenite phase.\cite{Buschbeck2009} This is in accordance to the XRD measurements of this film and confirms a mixed state at RT [Fig.~\ref{figure:xrd}]. The topography of the film originates from the twinning periodicity and the twinning angles of the involved variants. Thus, the surface angle $\alpha$ can be used to determine the involved twin structure.\cite{Buschbeck2009} From Fig.~\ref{figure:afm}(c), $\alpha=11^{\circ}$ is determined and the structure can be identified using the relation $c/a=\tan(45^{\circ}-\alpha/2)$. So, for Al-13.8 the topography leads to $c/a=0.84$, which roughly coincides with $c_{\text{14M}}/a_{\text{14M}}=0.89$ obtained from XRD analysis (Table~\ref{tab:latticeparameters}). A similar structure was also found for martensitic epitaxial Ni-Mn-Ga films.\cite{Buschbeck2009, Kaufmann2011}

The AFM micrograph of the completely martensitic film Al-12.7 in Fig.~\ref{figure:afm}(d) reveals two types of martensitic microstructure: Type A with traces inclined by 45$^{\circ}$ and periodicity of 88\,nm, and type B, which is almost flat and oriented parallel to the substrate etches. Type A is very similar to the microstructure of Al-13.8 [Fig.~\ref{figure:afm}(a)]. From the topography we can extract $\alpha=10^{\circ}$ leading to $c/a=0.84$ which agrees with $c_{\text{14M}}/a_{\text{14M}}=0.82$.
The type B microstructure shows shallow surface angles of about $1^{\circ}$ and could not doubtlessly be assigned to a certain twinning structure. This microstructure can be the origin of the additional reflections in the (004)$_{\text{NM}}$ pole figure of Al-12.7 in Fig.~\ref{figure:polefigures}(b), since it was not found in Al-13.8. 
Kaufmann et al. observe a similar type of topography and suggest that it originates from macroscopic NM variants.\cite{Kaufmann2011}

\subsection{Magnetic and Magnetocaloric Properties}
\label{sec:magn}
\begin{figure}
\centering
\includegraphics[width=7.7 cm]{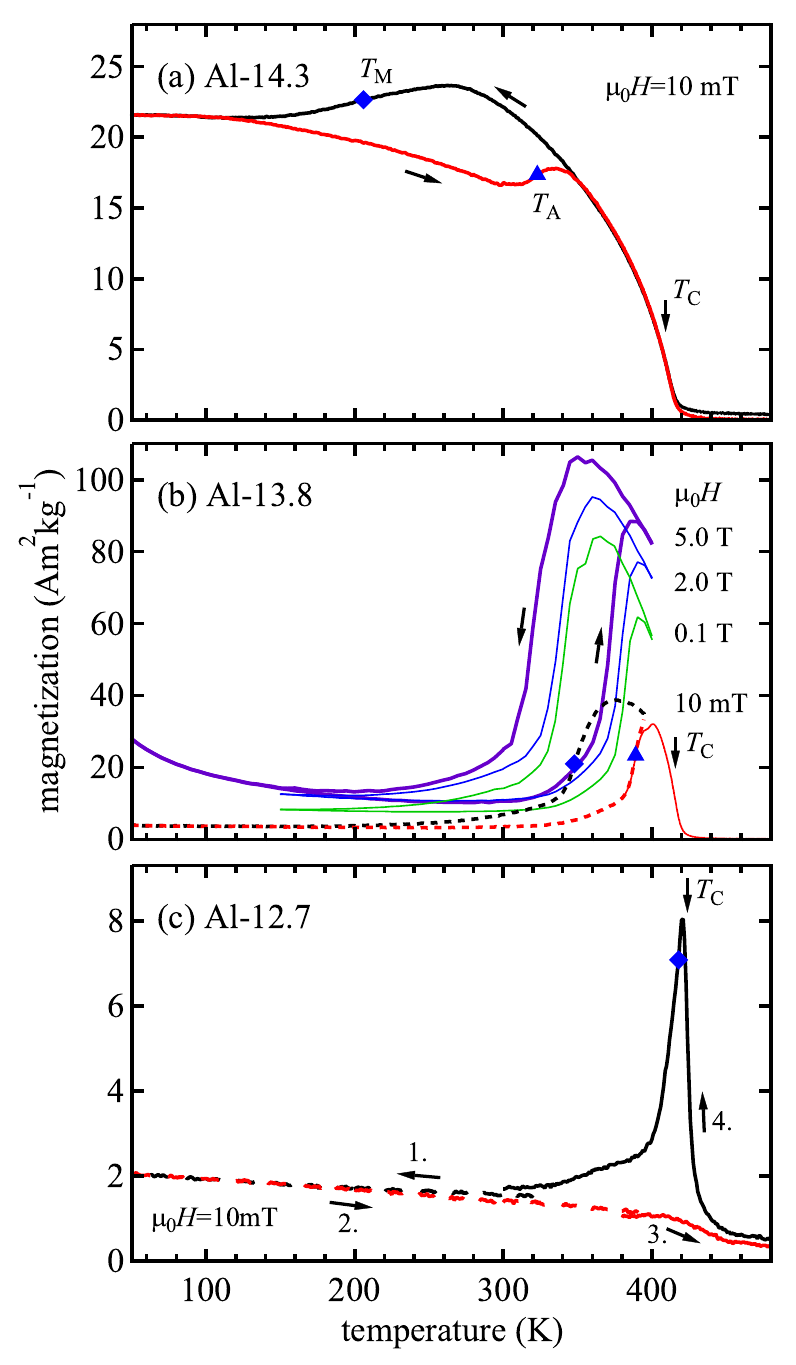}
\caption{\label{figure:isofield} $M(T)$ curves of a) A.-14.3, b) Al-13.8, and c) Al-12.7 for field cooling (black) and field heating (red) at $\mu_0H=10\,\text{mT}$. The martensitic transformation of Al-14.3 is incomplete resulting in a thermal hysteresis between 150\,K and 330\,K accompanied by only small changes of the magnetization.
For Al-13.8 $M(T)$ was measured at various fields up to 5\,T in order to estimate the field induced shift of the transformation temperatures and the magnetocaloric properties.
The blue squares and triangles depict $T_\text{M}$ and $T_\text{A}$ defined by the inflection points of the FC and FH curves. For each film $M(T)$ at $\mu_0H=10\,\text{mT}$ is presented as a combined curve where a cooling system and a heating system were employed indicated by dashed and solid lines, respectively, in b) and c). The measurement sequence and direction are indicated by numbers and arrows in c).}
\end{figure}

In order to analyze the metamagnetic characteristic of the martensitic transformation of Ni-Co-Mn-Al films, temperature dependent field cooling (FC) and field heating (FH) magnetization curves at 10\,mT applied field were measured and are shown in Fig.~\ref{figure:isofield}. Al-14.3 [Fig.~\ref{figure:isofield}(a)] shows a monotonically increasing magnetization below the austenite Curie temperature $T_{\text{C}}=409\,K$ with decreasing temperature down to 265\,K followed by a shallow decrease due to the martensitic transformation. The FC and FH curves envelop a thermal hysteresis between 150\,K and 330\,K.
However, the martensitic transformation of Al-14.3 is incomplete and residual austenite leads to high magnetization at low temperature. The large amount of residual austenite has been confirmed by low temperature XRD measurements (not shown). The transformation temperatures of the forward ($T_{\text{M}}$) and reverse ($T_{\text{A}}$) martensitic transformation and the austenite Curie temperature ($T_{\text{C}}$) were determined from the inflection points of the corresponding $M(T)$ curves.

For Al-13.8 [Fig.~\ref{figure:isofield}(b)], a distinct drop in the magnetization during cooling below 360\,K is visible, which is due to the magnetostructural transformation from a strongly ferromagnetic austenite to a weakly magnetic martensite.
The difference of magnetization between martensite and austenite, $\Delta M$, leads to a reduction of the transformation temperatures induced by the magnetic field. This follows from the magnetic Clausius-Clapeyron equation $\text{d}T_{\text{M}}/\text{d}H=-\Delta M/ \Delta S$ for two phases with entropy difference $\Delta S$.
Therefore, $M(H)$ curves at different applied fields up to 5\,T were measured [Fig.~\ref{figure:isofield}(b)].
For Al-13.8 $T_{\text{M}}=348\,\text{K}$ and $\text{d}T_{\text{M}}/\text{d}H\approx -4.3\,\text{K/T}$ for the forward transformation, and $T_{\text{A}}=389\,\text K$ and $\text{d}T_{\text{M}}/\text{d}H\approx -2.5\,\text{K/T}$ (between 0.1\,T and 5.0\,T) for the reverse transformation are obtained. The increase of magnetization below 100\,K at 5\,T is due to paramagnetic impurities in the MgO substrate. 

For bulk Ni$_{40}$Co$_{10}$Mn$_{33}$Al$_{17}$ $\text{d}T_{\text{M}}/\text{d}H\approx -6\,\text{K/T}$ and $\text{d}T_{\text{A}}/\text{d}H\approx -3.6\,\text{K/T}$ can be estimated from [\onlinecite{Kainuma2008}]. The reason for the higher values for bulk might be a larger $\Delta M$ because $M(T)$ drops nearly to zero below the martensitic transformation and the magnetization of the austenite phase is slightly higher.

Figure~\ref{figure:isofield}(c) shows the magnetization of Al-12.7. The FC curve depicts one sharp peak between $T_{\text{M}}$ and the close-by $T_{\text{C}}$.
In the FH curve $T_{\text{A}}$ is not visible in the magnetization because the reverse transformation occurs above $T_{\text{C}}$. Tab.~\ref{tab:samples} lists the Curie temperatures and martensitic transformation temperatures for all investigated films.
As observed for other Ni-Mn-based Heusler alloys $T_{\text{M}}$ is strongly dependent on the composition and increases with the valence electron concentration.\cite{Planes2009} $T_{\text{C}}$ shows a slight composition dependence, which can be explained by different Mn and Co concentrations. Both elements are known to increase $T_{\text{C}}$ with increasing concentration which is in agreement with our results (Table~\ref{tab:samples}). \cite{Cong2012, Wu2011, Mukadam2013}

The following analysis of the field induced reverse transformation and magnetocaloric properties is focused on the film Al-13.8 because it shows the largest $\Delta M$, which is the driving force for the field induced reverse transformation, and the martensitic transformation occurs close to room temperature, which is desired for magnetocaloric applications. 

\begin{figure}
\centering
\includegraphics[width=8 cm]{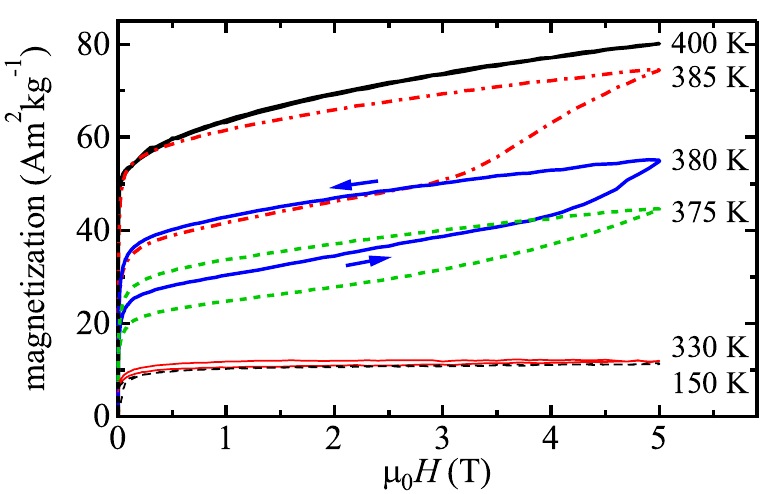}
\caption{\label{figure:isotherm} Isothermal magnetization measurements of Al-13.8 at different temperatures. Prior to each measurement the specimen was cooled to 150\,K. Around $T_{\text{A}}$ the magnetization shows significant hysteresis due to an irreversible field induced reverse transformation. The arrows indicate the direction of field change.}
\end{figure}

Figure~\ref{figure:isotherm} shows $M(H)_{T}$ isotherms at selected temperatures each measured after undercooling the specimen to 150\,K in order to assure well-defined starting conditions. In the martensitic phase (150\,K) the magnetization saturates below 1\,T whereas in the austenite phase (400\,K) up to 5\,T the magnetization does not saturate. The coercive field of the material is $H_{\text C}=1$\,mT at room temperature and thus the magnetic hysteresis caused thereby is negligible. However, at temperatures around $T_{\text{A}}$ the $M(H)_{T}$ show significant thermal hysteresis. During the initial increase of the external field beyond a critical value the slope of the $M(H)$ curve increases due to a field induced reverse transformation whereas under subsequent decreasing field the magnetization is consistently higher. During a second field loop the magnetization retraces the curve of decreasing field of the first loop (not shown). Hence the field induced reverse transformation is irreversible at the applied field of 5\,T, which is a consequence of the thermal hysteresis seen in Fig.~\ref{figure:isofield}(b).

In order to determine the magnetic entropy change $\Delta S_{\text{M}}(T)$ related to the martensitic transformation of Al-13.8 $M(T)_{H}$ FC and FH curves between 150\,K and 400\,K at external fields of 0.1\,T to 2.0\,T in steps of 0.1\,T, and 3\,T, 4\,T, and 5\,T have been measured. Figure~\ref{figure:isofield} shows a selection of these curves.
The magnetic entropy change under change of an applied field $\Delta H$ (from 0 to $H$) can be estimated by numerical evaluation of the integrated Maxwell relation 
\begin{equation}\label{eq:deltaS}
\Delta S_{\text{M}}(T,\Delta H)=\int_0^H \left(\frac{\partial M}{\partial T}\right)_{H} \text{d}H\text{.}
\end{equation}

\begin{figure}
\centering
\includegraphics[width=8 cm]{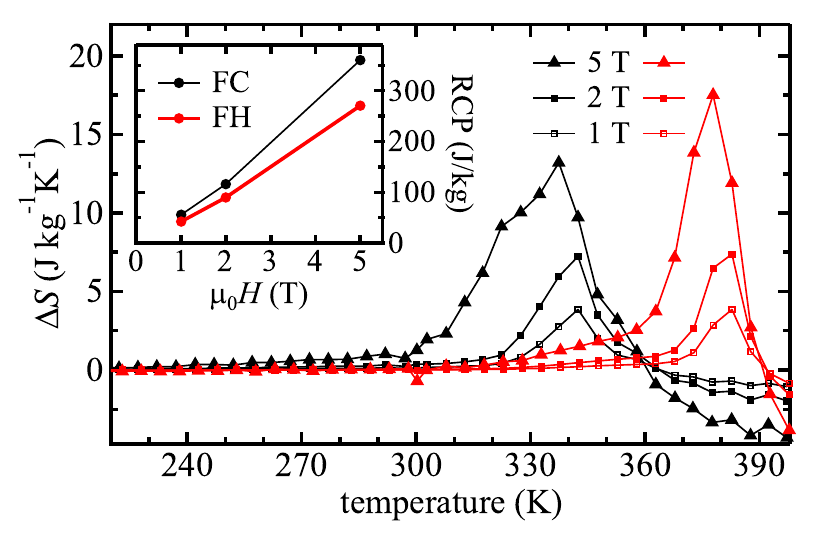}
\caption{\label{figure:magnetocaloric} Magnetic entropy change of Al-13.8 calculated from FC (black) and FH (red) $M(T)$ curves [Fig.~\ref{figure:isofield}(b)] using Eq.~\ref{eq:deltaS} for different magnetic field changes. The inset shows the change of the RCP with the external field.}
\end{figure}

Figure~\ref{figure:magnetocaloric} reveals $\Delta S_{\text{M}}(T)$ for different applied fields. Large values of 17.5\,J\,kg$^{-1}$K$^{-1}$ and 13.2\,J\,kg$^{-1}$K$^{-1}$ at a field change of $\mu_0 \Delta H=5\,\text{T}$ are obtained. For comparison Gd shows $\Delta S=-11\,\text{J}\,\text{kg}^{-1}\text{K}^{-1}$ for $\mu_0 \Delta H=5\,\text{T}$.\cite{Dan'kov1998}
An estimate of the full entropy change related to the martensitic transformation can be made using the Clausius-Clapeyron equation with the above given values for $\text{d}T_{\text{M}}/\text{d}H$ and $\text{d}T_{\text{A}}/\text{d}H$, and $\Delta M$=96\,Am$^2$kg$^{-1}$ for heating and 78\,Am$^2$kg$^{-1}$ for cooling. Therefore we obtain $\Delta S_{\text{M}}$=31\,J\,kg$^{-1}$K$^{-1}$ for heating and 22\,J\,kg$^{-1}$K$^{-1}$ for cooling.
Hence, a field change of 5\,T can induce an entropy change of up to 60\% of what expected for a full transformation.
Regarding the estimated values of the full entropy change the investigated Ni-Co-Mn-Al is comparable to Ni-Co-Mn-In.\cite{Kustov2009} 

For most of the other magnetocaloric Heusler compound thin films it is observed that the temperature range of the martensitic transformation is increased as compared to bulk, which results in broadening and flattening of the $\Delta S_{\text{M}}$ peak related to the martensitic transformation. For example epitaxial Ni-Co-Mn-In films only show $\Delta S=5\,\text{J}\,\text{kg}^{-1}\text{K}^{-1}$ for $\mu_0 \Delta H=6\,\text{T}$.\cite{Niemann2010} This broadening effect is less pronounced in Ni-Mn-Sn films\cite{Yuzuak2013} and the present Ni-Co-Mn-Al films. 
The reasons for the increase of the transformation range are not yet ascertained. Size effects, substrate clamping, and phase compatibility between martensite and austenite affect the characteristics of the martensitic transformation in thin films.\cite{Auge2012, Yuzuak2013} Also the heat treatment and thus the crystallization process during the preparation of thin films is completely different to bulk.

Therefore, the relative cooling power $\text{RCP}=\Delta S_{\text{M}}^{\text{max}}\delta T_{\text{FWHM}}$ is an appropriate measure to compare the magnetocaloric properties of bulk and thin film materials, where $\Delta S_{\text{M}}^{\text{max}}$ is the amplitude and $\delta T_{\text{FWHM}}$ the full width at half maximum of the corresponding peak. 
For the calculation we did not consider any losses due to hysteresis effects of the material which reduce the cooling efficiency under field cycling as suggested in [\onlinecite{Guillou2014}] and the RCP is primarily used as a measure for the area of the $\Delta S_{\text{M}}$ peak. However, it has to be stated that due to the thermal hysteresis the inverse MCE in the present material is irreversible using moderate magnetic fields without manually adjusting the temperature after each field cycle. Before the inverse MCE of the material can be utilized in an efficient cooling system it is necessary to tune the thermal hysteresis e.g. by optimization of the composition\cite{Teichert2015, Srivastava2010} or thermal treatment\cite{Ghosh2014}.

The inset of Fig.~\ref{figure:magnetocaloric} reveals the RCP of Al-13.8. The RCP calculated from FC curves is slightly larger than than from FH curves. This is due to the presence of the counteracting conventional magnetocaloric effect around $T_{\text C}$, which narrows the FH $\Delta S_{\text{M}}$ peak.
The promising magnetocaloric Heusler compound Ni-Co-Mn-In shows 19\,J\,kg$^{-1}$K$^{-1}$ at $\mu_0 \Delta H=1.9\,\text{T}$ for \textbf{bulk} Ni$_{45.7}$Co$_{5}$Mn$_{36.3}$In$_{13}$.\cite{Liu2012}
Otherwise, from [\onlinecite{Liu2012}] one can estimate the RCP of Ni-Co-Mn-In to 135\,J\,kg$^{-1}$ for $\mu_0 \Delta H=1.9\,\text{T}$ which is similar to the our results for $\mu_0 \Delta H=2\,\text{T}$.
For comparison, Gd shows $\text{RCP}=660\,\text{J\,kg}^{-1}$ for $\mu_0 \Delta H=5\,\text{T}$ (estimated from [\onlinecite{Dan'kov1998}]).
Also, the RCP of Ni-Co-Mn-Al is comparable to other promising thin film systems like FeRh\cite{Zhou2013}, MnAs\cite{Trassinelli2014} or La$_{0.8}$Ca$_{0.2}$MnO$_{3}$\cite{Debnath2013}, but in contrast to those only consists of common, inexpensive elements. 

\section{Conclusions}
In summary we have studied the structure, magnetism, and magnetocaloric properties of epitaxial Ni-Co-Mn-Al thin films on MgO substrates. The martensitic structure of these films was determined to be 14\textit{M}. We have introduced Ni-Co-Mn-Al as a magnetocaloric Heusler compound with large $\Delta S_{\text{M}}=17.5\,\text{J}\,\text{kg}^{-1}\text{K}^{-1}$ for Ni$_{41.0}$Co$_{10.4}$Mn$_{34.8}$Al$_{13.8}$ at a field change of $\mu_0 \Delta H=5\,\text{T}$. With the corresponding RCP of 271\,J\,kg$^{-1}$ it classifies as one of the most promising magnetocaloric thin film materials known so far. 

\begin{acknowledgments}
N. T., B. W. A. W., and A. H. gratefully acknowledge funding by the Deutsche Forschungsgemeinschaft through SPP 1599 ``Ferroic Cooling'' (Project Nos. HU 857/8-1, WA 3294/1-1). E. Y., I. D., and Y. E. thank The Scientific and Technological Research Council of Turkey (TUBITAK, Project No. 109T582) for funding. S.K. gratefully acknowledges funding by German Academic Exchange Service (DAAD, Project No. A/12/86233).
We also thank Kay Potzger from Helmholtz-Zentrum Dresden-Rossendorf for support regarding the MPMS measurements.
\end{acknowledgments}

\end{document}